\documentclass[aps,prl,twocolumn,showpacs,preprintnumbers,amsmath,superscriptaddress,amssymb,floats,nofootinbib]{revtex4}
\usepackage{graphicx}
\usepackage{hyperref}
\begin{document}


\title{Measurement of the proton electric to magnetic form factor ratio
from $\boldsymbol{^1\vec{\rm H}(\vec{\rm e},{\rm e}'{\rm p})}$}

\author{C.B.~Crawford}
\affiliation{Laboratory for Nuclear Science and Bates Linear Accelerator
  Center, Massachusetts Institute of Technology, Cambridge, MA 02139}  
\author{A.~Sindile}
\affiliation{University of New Hampshire, Durham, NH 03824}
\author{T.~Akdogan}
\affiliation{Laboratory for Nuclear Science and Bates Linear Accelerator
  Center, Massachusetts Institute of Technology, Cambridge, MA 02139} 
\author{R.~Alarcon}
\affiliation{Arizona State University, Tempe, AZ 85287}
\author{W.~Bertozzi}
\affiliation{Laboratory for Nuclear Science and Bates Linear Accelerator
  Center, Massachusetts Institute of Technology, Cambridge, MA 02139} 
\author{E.~Booth}
\affiliation{Boston University, Boston, MA 02215}
\author{T.~Botto}
\affiliation{Laboratory for Nuclear Science and Bates Linear Accelerator
  Center, Massachusetts Institute of Technology, Cambridge, MA 02139} 
\author{J.~Calarco}
\affiliation{University of New Hampshire, Durham, NH 03824}
\author{B.~Clasie}
\affiliation{Laboratory for Nuclear Science and Bates Linear Accelerator
  Center, Massachusetts Institute of Technology, Cambridge, MA 02139} 
\author{A.~DeGrush}
\affiliation{Laboratory for Nuclear Science and Bates Linear Accelerator
  Center, Massachusetts Institute of Technology, Cambridge, MA 02139} 
\author{T.W.~Donnelly}
\affiliation{Laboratory for Nuclear Science and Bates Linear Accelerator
  Center, Massachusetts Institute of Technology, Cambridge, MA 02139} 
\author{K.~Dow}
\affiliation{Laboratory for Nuclear Science and Bates Linear Accelerator
  Center, Massachusetts Institute of Technology, Cambridge, MA 02139} 
\author{D.~Dutta}
\affiliation{Duke University, Durham, NC 27708-0305}
\author{M.~Farkhondeh}
\affiliation{Laboratory for Nuclear Science and Bates Linear Accelerator
  Center, Massachusetts Institute of Technology, Cambridge, MA 02139} 
\author{R.~Fatemi}
\affiliation{Laboratory for Nuclear Science and Bates Linear Accelerator
  Center, Massachusetts Institute of Technology, Cambridge, MA 02139} 
\author{O.~Filoti}
\affiliation{University of New Hampshire, Durham, NH 03824}
\author{W.~Franklin}
\affiliation{Laboratory for Nuclear Science and Bates Linear Accelerator
  Center, Massachusetts Institute of Technology, Cambridge, MA 02139} 
\author{H.~Gao}
\affiliation{Duke University, Durham, NC 27708-0305}
\author{E.~Geis}
\affiliation{Arizona State University, Tempe, AZ 85287}
\author{S.~Gilad}
\affiliation{Laboratory for Nuclear Science and Bates Linear Accelerator
  Center, Massachusetts Institute of Technology, Cambridge, MA 02139} 
\author{W.~Haeberli}
\affiliation{University of Wisconsin, Madison, WI 53706}
\author{D.~Hasell}
\affiliation{Laboratory for Nuclear Science and Bates Linear Accelerator
  Center, Massachusetts Institute of Technology, Cambridge, MA 02139} 
\author{W.~Hersman}
\affiliation{University of New Hampshire, Durham, NH 03824}
\author{M.~Holtrop}
\affiliation{University of New Hampshire, Durham, NH 03824}
\author{P.~Karpius}
\affiliation{University of New Hampshire, Durham, NH 03824}
\author{M.~Kohl}
\affiliation{Laboratory for Nuclear Science and Bates Linear Accelerator
  Center, Massachusetts Institute of Technology, Cambridge, MA 02139} 
\author{H.~Kolster}
\affiliation{Laboratory for Nuclear Science and Bates Linear Accelerator
  Center, Massachusetts Institute of Technology, Cambridge, MA 02139} 
\author{T.~Lee}
\affiliation{University of New Hampshire, Durham, NH 03824}
\author{A.~Maschinot}
\affiliation{Laboratory for Nuclear Science and Bates Linear Accelerator
  Center, Massachusetts Institute of Technology, Cambridge, MA 02139} 
\author{J.~Matthews}
\affiliation{Laboratory for Nuclear Science and Bates Linear Accelerator
  Center, Massachusetts Institute of Technology, Cambridge, MA 02139} 
\author{K.~McIlhany}
\affiliation{United States Naval Academy, Annapolis, MD 21402}
\author{N.~Meitanis}
\affiliation{Laboratory for Nuclear Science and Bates Linear Accelerator
  Center, Massachusetts Institute of Technology, Cambridge, MA 02139} 
\author{R.G.~Milner}
\affiliation{Laboratory for Nuclear Science and Bates Linear Accelerator
  Center, Massachusetts Institute of Technology, Cambridge, MA 02139} 
\author{J.~Rapaport}
\affiliation{Ohio University, Athens, OH 45701}
\author{R.P.~Redwine}
\affiliation{Laboratory for Nuclear Science and Bates Linear Accelerator
  Center, Massachusetts Institute of Technology, Cambridge, MA 02139} 
\author{J.~Seely}
\affiliation{Laboratory for Nuclear Science and Bates Linear Accelerator
  Center, Massachusetts Institute of Technology, Cambridge, MA 02139} 
\author{A.~Shinozaki}
\affiliation{Laboratory for Nuclear Science and Bates Linear Accelerator
  Center, Massachusetts Institute of Technology, Cambridge, MA 02139} 
\author{S.~\v{S}irca}
\affiliation{Laboratory for Nuclear Science and Bates Linear Accelerator
  Center, Massachusetts Institute of Technology, Cambridge, MA 02139} 
\author{E.~Six}
\affiliation{Arizona State University, Tempe, AZ 85287}
\author{T.~Smith}
\affiliation{Dartmouth College, Hanover, NH 03755}
\author{B.~Tonguc}
\affiliation{Arizona State University, Tempe, AZ 85287}
\author{C.~Tschalaer}
\affiliation{Laboratory for Nuclear Science and Bates Linear Accelerator
  Center, Massachusetts Institute of Technology, Cambridge, MA 02139} 
\author{E.~Tsentalovich}
\affiliation{Laboratory for Nuclear Science and Bates Linear Accelerator
  Center, Massachusetts Institute of Technology, Cambridge, MA 02139} 
\author{W.~Turchinetz}
\affiliation{Laboratory for Nuclear Science and Bates Linear Accelerator
  Center, Massachusetts Institute of Technology, Cambridge, MA 02139} 
\author{J.F.J.~van~den~Brand}
\affiliation{Vrije Universitaet and NIKHEF, Amsterdam, The Netherlands}
\author{J.~van~der~Laan}
\affiliation{Laboratory for Nuclear Science and Bates Linear Accelerator
  Center, Massachusetts Institute of Technology, Cambridge, MA 02139} 
\author{F.~Wang}
\affiliation{Laboratory for Nuclear Science and Bates Linear Accelerator
  Center, Massachusetts Institute of Technology, Cambridge, MA 02139} 
\author{T.~Wise}
\affiliation{University of Wisconsin, Madison, WI 53706}
\author{Y.~Xiao}
\affiliation{Laboratory for Nuclear Science and Bates Linear Accelerator
  Center, Massachusetts Institute of Technology, Cambridge, MA 02139} 
\author{W.~Xu}
\affiliation{Duke University, Durham, NC 27708-0305}
\author{C.~Zhang}
\affiliation{Laboratory for Nuclear Science and Bates Linear Accelerator
  Center, Massachusetts Institute of Technology, Cambridge, MA 02139} 
\author{Z.~Zhou}
\affiliation{Laboratory for Nuclear Science and Bates Linear Accelerator
  Center, Massachusetts Institute of Technology, Cambridge, MA 02139} 
\author{V.~Ziskin}
\affiliation{Laboratory for Nuclear Science and Bates Linear Accelerator
  Center, Massachusetts Institute of Technology, Cambridge, MA 02139} 
\author{T.~Zwart}
\affiliation{Laboratory for Nuclear Science and Bates Linear Accelerator
  Center, Massachusetts Institute of Technology, Cambridge, MA 02139} 

\noaffiliation

\date{\today}

\begin{abstract}
  We report the first precision measurement of the proton electric to magnetic
  form factor ratio from spin-dependent elastic scattering of longitudinally
  polarized electrons from a polarized hydrogen internal gas target.  The
  measurement was performed at the MIT-Bates South Hall Ring over a
  range of four-momentum transfer squared $Q^2$ from 0.15 to 0.65 (GeV/c)$^2$.
  Significantly improved results on the proton electric and magnetic form
  factors are obtained in combination with previous cross-section data on
  elastic electron-proton scattering in the same $Q^2$ region.

\end{abstract}


\maketitle

The electromagnetic form factors of the nucleon are fundamental quantities
describing the distribution of charge and magnetization within the nucleon.
At low four-momentum transfer squared $Q^2$, they are sensitive to the pion
cloud~\cite{friedrich,faessler}, and provide tests of effective field theories
of quantum chromodynamics (QCD) based on chiral symmetry~\cite{schindler}.
Lattice QCD has made considerable progress in describing the form factors at
low $Q^2$~\cite{lattice} and, with future advances in technique and computing
power, tests against precise data will be possible.  Accurate measurements of
nucleon electromagnetic form factors at low $Q^2$ are also important for
interpretation of parity-violation electron scattering experiments~\cite{g0},
which probe the strange quark contribution to the nucleon electromagnetic
structure.  Knowledge of the internal structure of protons and neutrons in
terms of the quark and gluon degrees of freedom of QCD
provides a basis for understanding more complex, strongly interacting matter
at the level of quarks and gluons.

The proton electric ($G_E^p$) and magnetic ($G_M^p$) form factors have been
studied extensively in the past \cite{gao} over a wide range of $Q^2$ from
unpolarized electron-proton (e-p) elastic scattering using the Rosenbluth (L-T)
separation technique \cite{rosenbluth}.  While the precise knowledge of
the separated form factors $G_E^p$ and $G_M^p$ is important for understanding
the underlying electromagnetic structure of the nucleon, it is also very
interesting to study the ratio $\mu_pG_E^p/G_M^p$ as a function of $Q^2$,
where $\mu_p\sim 2.79$ is the proton magnetic moment in units of nuclear
magnetons.  The observation of a $Q^2$ dependence in the form factor ratio
would suggest different charge and current spatial distributions inside the
proton. 

Recent advances in polarized beams, targets, and polarimetry have made
possible a new class of experiments extracting $\mu_pG_E^p/G_M^p$ utilizing
spin degrees of freedom. Extraction of the form factor ratio from double
polarization observables has two substantial advantages over unpolarized cross
section measurements.  First, the spin-dependent cross section has an
interference term between $G_E^p$ and $G_M^p$, allowing for a direct
determination of $\mu_pG_E^p/G_M^p$ from either the spin-dependent
asymmetry~\cite{donnelly} or the recoil polarization
measurement~\cite{arnold}, whereas the unpolarized cross section depends only
on the squares of $G_E^p$ and $G_M^p$.  Second, spin degrees of freedom can be
varied instead of the beam energy and scattering angle (as is done in the
Rosenbluth separation), greatly reducing the systematic errors.  New data from
polarization transfer experiments \cite{mjones,gayou} show a very
intriguing behavior at higher $Q^2$: starting at $Q^2=1~({\rm GeV/c})^2$,
$\mu_pG_E^p/G_M^p$ drops linearly from approximately 1 down to 0.28 at the
highest measured $Q^2$ value ($\sim$ 5.54 (GeV/c)$^2$). This is very different
from unity, as suggested by previous unpolarized cross section
measurements~\cite{walker,andivahis} and verified by recent
experiments~\cite{hallc,arrington}.


While the interesting $Q^2$ dependence of the proton form factor ratio 
from recoil polarization experiments~\cite{mjones,gayou} has been
described in terms of nonzero parton orbital angular momentum or 
hadron helicity
flip~\cite{belitsky,ralston02,miller,brodsky_ff}, it is important to
understand the discrepancy between results obtained from recoil proton
polarization measurements and those from the Rosenbluth method. 
Two-photon exchange contributions may contribute~\cite{wally,tpe_theory} 
up to about half of the observed discrepancy between the two experimental
methods, identifying the need for further experimental tests.

In this letter we report the first measurement of $\mu_pG_E^p/G_M^p$ from
$\vec{^1\rm H}(\vec{\rm e},{\rm e}'{\rm p})$ in the $Q^2$ region between 0.15
and 0.65 (GeV/c)$^2$~\cite{thesis_cc,thesis_as}, overlapping with the lower
$Q^2$ region of the recoil polarization
data~\cite{milbrath,dieterich,pospischil,mjones,gayou1}.  This is an important
region which allows for tests of effective field theory predictions and
verifications of future precision results of Lattice QCD.  It also helps to
quantify the role of the pion cloud in the structure of the nucleon.  The
polarized target technique has different sources of systematic uncertainty
than recoil polarimetry, but still benefits from the same cancellations in
systematic uncertainties such as detection efficiency and luminosity.

In the one-photon exchange approximation, the elastic scattering asymmetry of
longitudinally polarized electrons from polarized protons with respect to the
electron beam helicity has the form~\cite{donnelly}
\begin{equation}
  A_{phys} = \frac
  { v_z \cos\theta^* {G_M^p}^2 +
    v_x \sin\theta^* \cos\phi^* G_M^p G_E^p }
  { (\tau {G_M^p}^2 + \epsilon {G_E^p}^2)
    \; / \; [\epsilon(1+\tau)] },
  \label{eq:asym}
\end{equation}
where $\theta^*$ and $\phi^*$ are
the polar and azimuthal angles of the target polarization defined relative to
the three-momentum transfer vector of the virtual photon and
$\tau=Q^2/(4M_p^2)$ with the proton mass $M_p$.  The longitudinal polarization
of the virtual photon is denoted as
$\epsilon=[1+2(1+\tau)\tan^2(\theta_e/2)]^{-1}$ where $\theta_e$ is the
electron scattering angle, and $v_z = -2\tau \tan(\theta_e/2)
\sqrt{1/(1+\tau)+\tan^2(\theta_e/2)}$, $v_x = -2\tan(\theta_e/2)
\sqrt{\tau/(1+\tau)}$ are kinematic factors.  The experimental asymmetry
\begin{equation}
A_{exp} = P_{b} P_{t}\, A_{phys}
\label{eq:exp_asym}
\end{equation} 
is reduced by the beam ($P_{b}$) and target ($P_{t}$) polarizations.


The form factor ratio $\mu_p G_E^p/G_M^p$ and the polarization product
$P_{b}P_{t}$ can be determined by measuring two experimental asymmetries $A_l$
and $A_r$ at the same $Q^2$ value, but with different spin orientations
$(\theta_l^*,\phi_l^*)$ and $(\theta_r^*,\phi_r^*)$, respectively.  For a
detector configuration that is symmetric about the incident electron beam and
with the target polarization angle oriented $\sim 45^\circ$ to the left of the
beam, $A_l$ and $A_r$ can be measured simultaneously by forming two
independent asymmetries of electrons scattered into the beam-left and
beam-right sectors, respectively.  In this case $A_l$ ($A_r$) is predominantly
transverse (longitudinal).

The Bates Large Acceptance Spectrometer Toroid (BLAST) experiment was carried
out in the South Hall Ring of the MIT Bates Linear Accelerator Center, which
stored an intense polarized beam with a beam current of up to 225~mA and
longitudinal electron polarization of 65\%.  A 180$^\circ$ spin rotator
(Siberian Snake) was used in the ring opposite the interaction point to
preserve the longitudinal electron polarization at the target, which was
continuously monitored with a Compton polarimeter installed upstream of the
internal target region.  The ring was filled every fifteen minutes,
alternating electron helicity on successive fills.

The electrons scattered from polarized protons in a cylindrical, windowless
aluminum target tube 60~cm long by 15~mm in diameter.  The polarized protons
were fed from an Atomic Beam Source (ABS) located above the target, well
shielded against the strong, spatially varying magnetic field of the
toroid~\cite{abs}.  A 10 mm diameter tungsten collimator in front of the
target protected the cell wall coating from exposure to the beam and minimized
the background rate in the detector.  The ABS provided highly polarized
($P_t\sim 80\%$) isotopically pure hydrogen atoms.  The spin state was
randomly changed every five minutes by switching the final RF transition
before the target to ensure equal target intensities for both states.  The
average target spin direction was oriented 48.0$^\circ$ to the left of the
beam direction using a 0.04~T holding field.


The achieved luminosity ${\mathcal L} = 4.4 \times 10^{31}~{\rm cm}^{-2}{\rm
  s}^{-1}$ of the internal gas target required the use of a large acceptance
spectrometer.  The symmetric detector package was built around eight copper
coils which provided a maximum 0.38~T toroidal magnetic field at 6730 A,
resulting in an integrated field strength of 0.15--0.44~Tm within the drift
chambers for momentum analysis.  Two of the eight sectors covering scattering
angles of 23$^\circ$--76$^\circ$ and $\pm 15^\circ$ out of plane were
instrumented with: three drift chambers each for momentum, angle, and position
determination of charged tracks, plastic scintillators for triggering and
time-of-flight particle identification, and \v Cerenkov detectors for pion
rejection.  Details of the BLAST detector can be found in~\cite{hasell}.

Data were acquired for 89.8~pb$^{-1}$ of integrated luminosity, corresponding
to 298~kC of integrated charge on the target.  The elastic events were
detected in coincidence with a hardware trigger requirement of scintillator
signals for both the electron and proton.  A second-level trigger additionally
required signals in the wire chambers to reduce excessive trigger rates and to
decrease the computer deadtime.  The beam current was measured using a
parametric direct current transformer in the ring, gated by the DAQ deadtime.

The elastic events were selected with a cut on the invariant mass of the
virtual photon and the target proton system, fiducial cuts on the polar and
azimuthal acceptance, and cuts on the position of the electron and proton
vertex in the target cell.  These cuts were consistent with kinematic cuts on
angle, momentum and timing correlations between the scattered electron and the
recoil proton, made possible by the overdetermination of the elastic reaction.
The cuts were sufficient to reduce the background to less than 1.5\% without
decreasing the elastic yield.  The remaining background was measured with
14.9~kC of integrated charge on the same target cell without hydrogen flowing.
The dependence of the background rates from the target cell wall on the target
density due to a widening of the beam diameter (halo effect) was found to be
negligible by comparing the quasielastic (e,e$'$n) rates between hydrogen and
empty targets.

Separate yields $\sigma_{ij}$ were analyzed for each combination of electron
helicity $i$ and target spin state $j$, normalized by the integrated beam
current.  They were divided into eight $Q^2$ bins, listed in
Table~\ref{tab:ffr}.  The event-weighted $\langle Q^2 \rangle$ was formed from
the average of $\langle Q^2_e\rangle$ (determined from the electron scattering
angle) and $\langle Q^2_p\rangle$ (from the proton recoil angle) in each bin.
The yield distributions were in good agreement with results from a Monte Carlo
simulation, including all detector efficiencies. 

The experimental double asymmetry was formed from $(\sigma_{++} - \sigma_{+-}
- \sigma_{-+} + \sigma_{--}) / \Sigma\sigma_{ij}$.  The beam and target
single-spin asymmetries were also analyzed and served as a monitor of false
asymmetries, which were found to be negligible.  The experimental asymmetry
was corrected for dilution by unpolarized background, including the beam halo
effect. Radiative corrections were also applied using the code
MASCARAD~\cite{afanasev}, but were less than 0.43\% for $A_r$ and 0.16\% for
$A_l$.  The measured spin-dependent physics asymmetries are shown in Fig.~1
(upper panel), with $P_{b}P_{t}$ determined from the model-independent
analysis described below, and are compared with the calculated asymmetries
based on the parameterization for the proton form factors by H\"{o}hler {\it
  et al.}~\cite{hohler}.

\begin{figure}
  \centering
  \includegraphics[width=.48\textwidth]{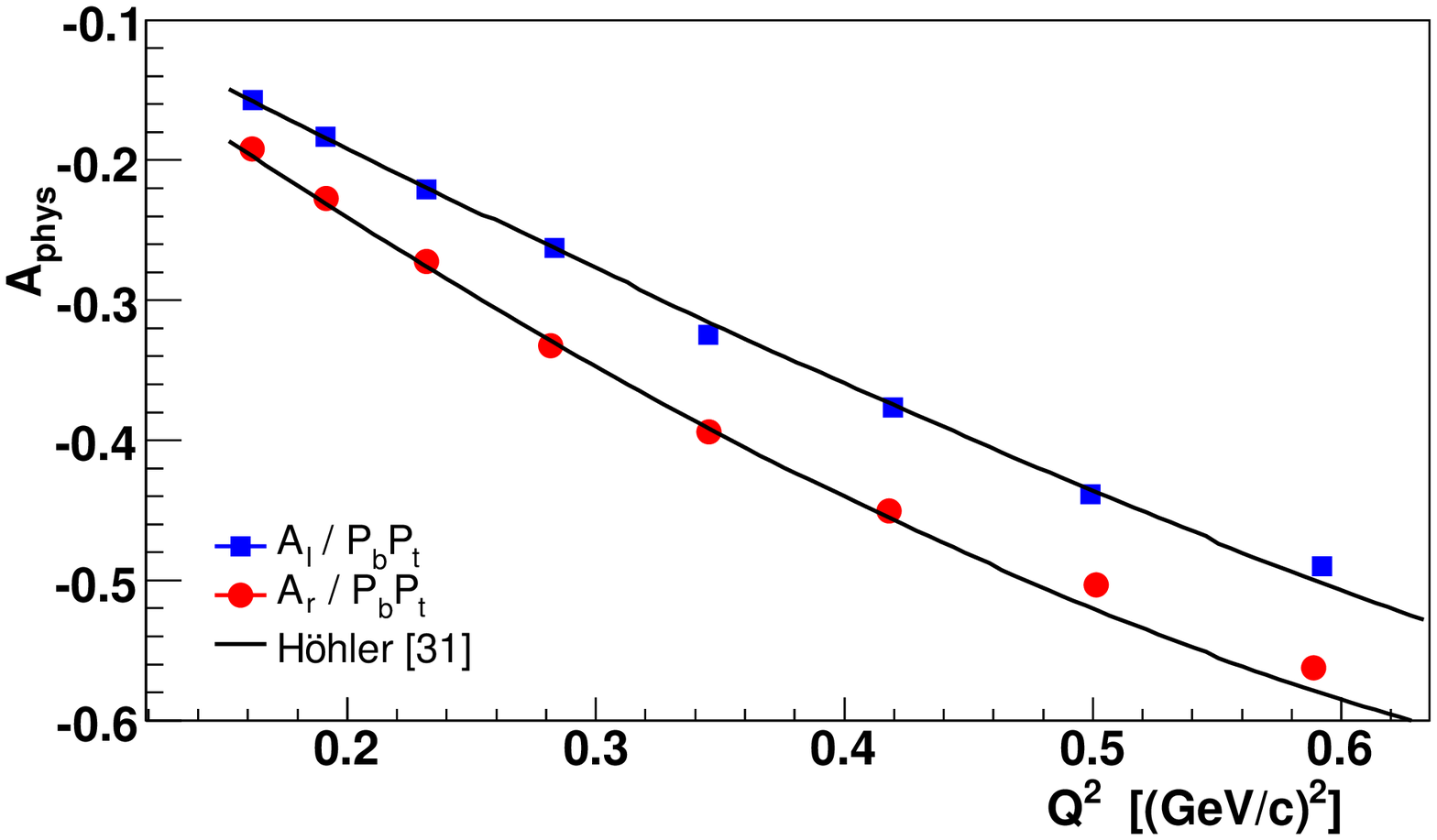}
  \includegraphics[width=.48\textwidth]{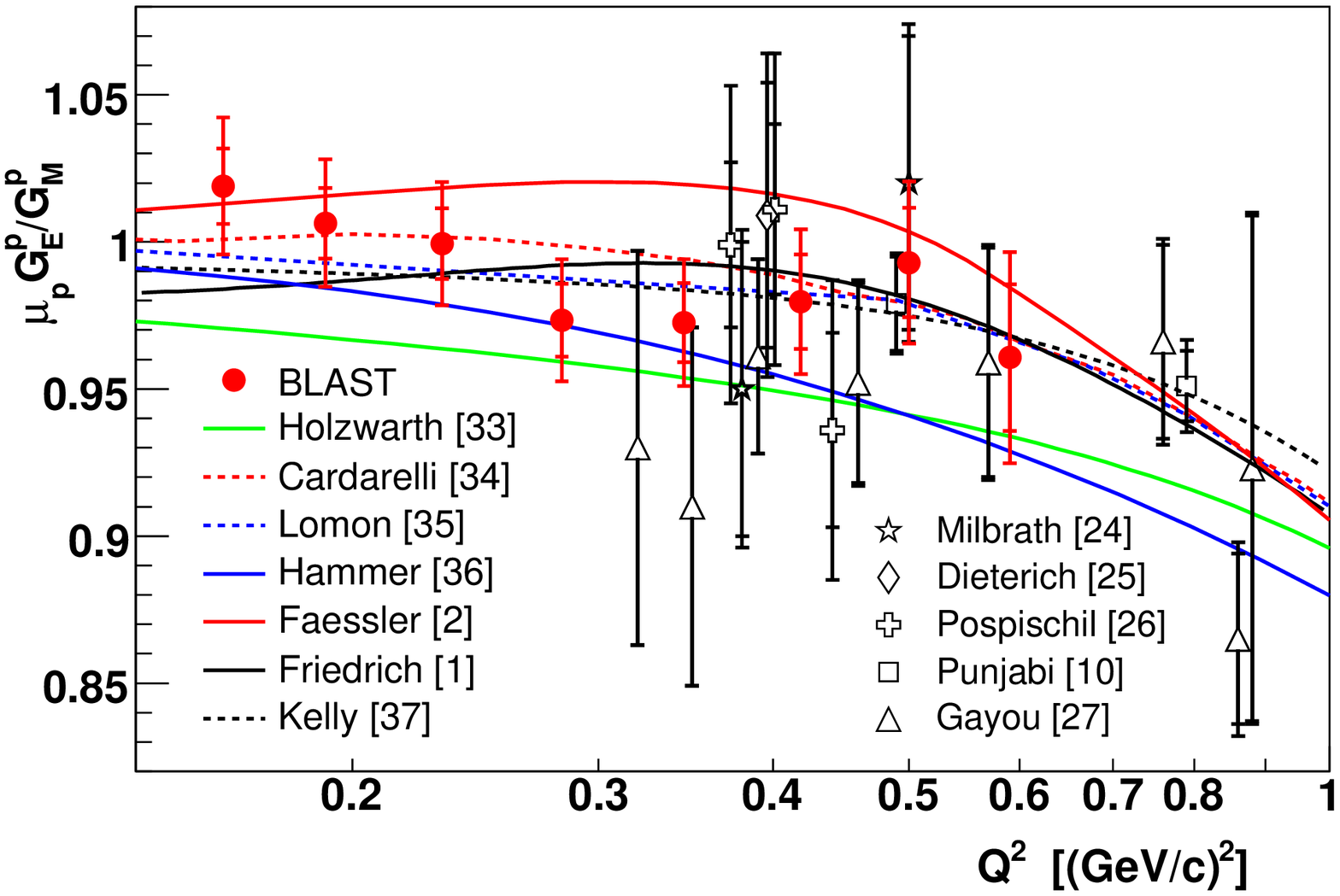}
  \caption{
    Upper panel: Spin-dependent $\vec{^1{\rm H}}(\vec{\rm e},{\rm e}'{\rm p})$ 
    asymmetry ($P_bP_t = 0.537 \pm 0.003$) compared to the expected asymmetries
    based on the parameterization~\cite{hohler} for the nucleon form factors.  
    Lower panel: Results of $\mu_pG_E^p/G_M^p$ shown with the world
    polarized data~\cite{milbrath,dieterich,pospischil,mjones,gayou1} and several
    models~\cite{holzwarth1,simula3,lomon2,hammer,faessler,friedrich,kelly}
    described in the text.}
  \label{fig:ffr}
\end{figure}

To extract the form factor ratio, the experimental asymmetries $A_l$ and $A_r$
were interpolated in each $Q^2$ bin to the average value of $\langle Q^2
\rangle$ in the left and right sectors.  As discussed previously, the
polarization product $P_bP_t$ and the form factor ratio $\mu_pG_E^p/G_M^p$ can
be determined from the measured asymmetries $A_{l}$ and $A_{r}$ using
Eqs.~(\ref{eq:asym}), (\ref{eq:exp_asym}).  This way the so-called super ratio
$A_{l} / A_{r}$ would yield $\mu_pG_E^p/G_M^p$ independent of $P_bP_t$ for
each $Q^2$ bin. In our final analysis, however, a single value of $P_bP_t$ was
fit for all $Q^2$ values for optimal extraction of the form factor
ratio~\cite{thesis_cc}, resulting in $P_bP_t = 0.537 \;\pm\; 0.003$~(stat)
$\pm\; 0.007$~(sys). 

The dominant source of systematic uncertainty was the determination of
$\langle Q^2 \rangle$, estimated from the difference between $\langle Q^2_e
\rangle$ and $\langle Q^2_p \rangle$ to be less than 0.002~(GeV/c)$^2$ in each
bin and varying from point to point.  The event-weighted average spin angle of
the target with respect to the beam was $48.0^\circ \pm 0.5^\circ$, extracted
from the analysis of the $T_{20}$ tensor analyzing power in elastic scattering
from deuterium in combination with a \squeezetable
\begin{table}[ht]
  \centering
  \begin{tabular}{r@{$-$}l@{\quad}c@{\quad}c@{(}c@{)\quad}c@{(}c@{)\quad}c@{$\pm$}c@{$\pm$}c}
    \hline\hline
    \multicolumn{2}{c@{\qquad}}{$Q^2$ bin} & 
    \multicolumn{1}{c@{\quad}}{$\langle Q^2 \rangle$} & 
    \multicolumn{2}{c@{\quad}}{$A_l$} & 
    \multicolumn{2}{c@{\quad}}{$A_r$} & 
    \multicolumn{3}{c@{\qquad}}{$\mu_pG_E^p/G_M^p$} \\ 
    \hline
    0.150 & 0.175 & 0.162 & -0.0838 & 15 & -0.1022 & 13 & 1.019 & 0.013 & 0.015 \\
    0.175 & 0.211 & 0.191 & -0.0976 & 14 & -0.1213 & 14 & 1.006 & 0.012 & 0.014 \\
    0.211 & 0.257 & 0.232 & -0.1178 & 17 & -0.1453 & 17 & 0.999 & 0.012 & 0.012 \\
    0.257 & 0.314 & 0.282 & -0.1403 & 22 & -0.1768 & 20 & 0.973 & 0.012 & 0.011 \\
    0.314 & 0.382 & 0.345 & -0.1730 & 26 & -0.2100 & 25 & 0.973 & 0.014 & 0.010 \\
    0.382 & 0.461 & 0.419 & -0.2011 & 31 & -0.2397 & 33 & 0.980 & 0.016 & 0.009 \\
    0.461 & 0.550 & 0.500 & -0.2333 & 39 & -0.2686 & 40 & 0.993 & 0.019 & 0.008 \\
    0.550 & 0.650 & 0.591 & -0.2618 & 54 & -0.2994 & 57 & 0.961 & 0.025 & 0.007 \\
    \hline\hline
  \end{tabular}
  \caption{Results of measured experimental elastic asymmetries $A_l, A_r$
    (with statistical uncertainties)
    and extracted proton form factor ratio
    $\mu_pG_E^p/G_M^p \pm \rm{stat.} \pm \rm{sys.}$ uncertainties. The values
    for $Q^2$ are given in (GeV/c)$^2$.}  
  \label{tab:ffr}
\end{table}
careful mapping of the magnetic field in the target region~\cite{zhang}.
However, the proton form factor ratio has reduced sensitivity to the target
spin angle uncertainty due to a compensation in the simultaneous extraction of
$P_bP_t$.

\begin{figure}
\centering
  \includegraphics[width=.48\textwidth]{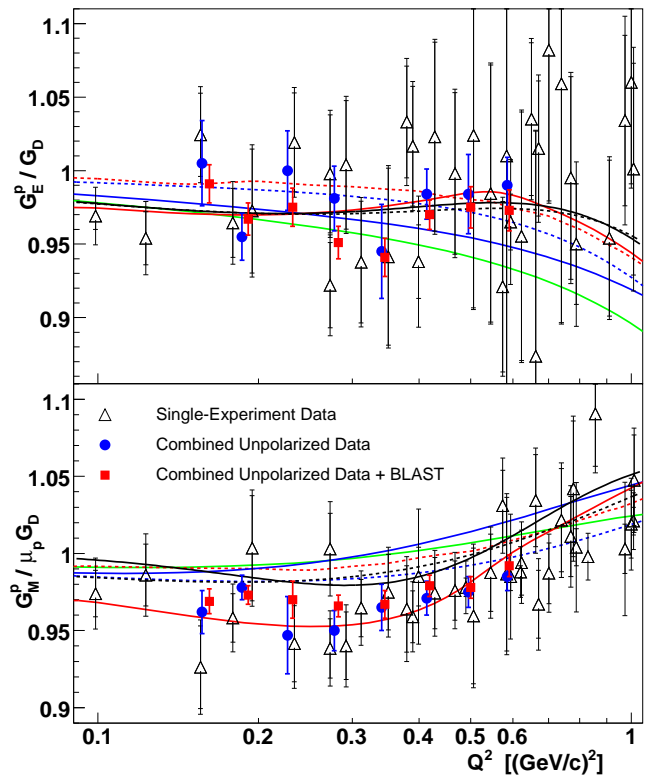}
  \caption{Compilation of world data on $G_E^p/G_D$ and $G_M^p/\mu_pG_D$ at
    BLAST kinematics with (red) and without (blue) BLAST input, shown with
    total uncertainties.  The results from single-experiment L-T
    separations~\cite{walker,hallc,janssens,price,berger,bartel73,borkowski,bosted}
    are shown with open symbols.  The curves have the same meaning as in
    Fig.~\ref{fig:ffr}.}
  \label{fig:blast_ff}
\end{figure}

The results are listed in Table~\ref{tab:ffr} and are displayed in
Fig.~\ref{fig:ffr} with the inner error bars due to statistical uncertainties
and the outer error bars being the total (statistical and systematic
contributions added in quadrature).  Also shown in Fig.~\ref{fig:ffr} are
published recoil polarization
data~\cite{milbrath,dieterich,pospischil,mjones,gayou1}, together with a few
selected models: a soliton model~\cite{holzwarth1}, 
a relativistic constituent quark model (CQM) with SU(6) symmetry breaking and
a constituent quark form factor~\cite{simula3}, an extended vector meson
dominance model~\cite{lomon2}, an updated dispersion model~\cite{hammer}, and
a Lorentz covariant chiral quark model~\cite{faessler}.  We also show the
parameterizations by Friedrich and Walcher~\cite{friedrich} and
Kelly~
The impact of the BLAST results oLAST results on the separated proton charge and magnetic
form factors normalized to the dipole form factor $G_D = (1 +
Q^2/0.71)^{-2}$ is illustrated in Fig.~\ref{fig:blast_ff}.  In this figure,
Rosenbluth extractions of $G_E^p$ and $G_M^p$ from single
experiments~\cite{walker,hallc,janssens,price,berger,bartel73,borkowski,bosted}
are presented as open triangles with statistical and total error bars, the
systematic errors added in quadrature.  The combined cross section
data~\cite{hallc,janssens,price,berger,borkowski,bosted,bartel66,stein,niculescu},
obtained from \cite{arrington69,borkowski}, were binned according to
Table~\ref{tab:ffr} to obtain a single L-T separation of $G_E^p$ and $G_M^p$
at each of the BLAST kinematics (blue circles).  In comparison, the red
squares show the form factors extracted from the combination of unpolarized
cross sections and the measured form factor ratio from BLAST.  Not only are
the uncertainties reduced by a factor of 1.3--2.5, but also the negative
correlation between $G_E^p$ and $G_M^p$ typical of L-T separations is greatly
reduced.  Details of the extraction will follow in a separate paper.

The proton form factors suggest a rather interesting structure around $Q^2
\approx 0.2$ to 0.5 (GeV/c)$^2$.  While the ratio of the electric and magnetic
form factors in this region is consistent with unity, the separated form
factors show a deviation from the leading dependence given by the dipole form
factor below 1~(GeV/c)$^2$. Similar structure has been observed in the neutron
electric~\cite{gen} and magnetic form factor~\cite{xu} data as well.  A
possible explanation for this effect could be due to a manifestation of the
pion cloud at low momentum transfer~\cite{friedrich,faessler}, though a more
detailed understanding of this structure is expected from chiral models and
ultimately from Lattice QCD.

We thank the staff at the MIT-Bates Linear Accelerator Center for the delivery
of high quality electron beam and for their technical support.  This work is
supported in part by the U.S. Department of Energy and the National Science
Foundation.

\end{document}